%% file: paper.tex
\documentclass[10pt,letterpaper]{article}

\include{macros}

\usepackage{hyperref}
\hypersetup{
  colorlinks=true,
  linkcolor=blue,
  citecolor=blue,
  urlcolor=blue,
  breaklinks=true}
\usepackage{ifpdf}
\ifpdf
\fi
\usepackage{cite}
\begin{document}

\title{High-bandwidth transfer of phase stability through a fiber frequency comb}

\author{%
  Nils~Scharnhorst,\textsuperscript{1,3}
  Jannes~B.~Wübbena,\textsuperscript{1,3}
  Stephan~Hannig,\textsuperscript{1}
  Kornelius~Jakobsen,\textsuperscript{1}
  Johannes~Kramer,\textsuperscript{1}
  Ian~D.~Leroux,\textsuperscript{1*}
  and Piet~O.~Schmidt\textsuperscript{1,2}
}
\address{%
  \textsuperscript{1}
  QUEST Institut für Experimentelle Quantenmetrologie,
  Physikalisch-Technische Bundesanstalt,
  38116 Braunschweig, Germany \\
  \textsuperscript{2}
  Institut für Quantenoptik,
  Leibniz Universität Hannover,
  30167 Hannover, Germany \\
  \textsuperscript{3}
  These authors contributed equally.
}
\email{\textsuperscript{*}idleroux@quantummetrology.de}

\begin{abstract}
We demonstrate phase locking of a \SI{729}{\nm} diode laser to a \SI{1542}{\nm} master laser via an erbium-doped-fiber frequency comb,
using a transfer-oscillator feedforward scheme which suppresses the effect of comb noise in an unprecedented \SI{1.8}{\MHz} bandwidth.
We illustrate its performance by carrying out coherent manipulations of a trapped calcium ion with \SI{99}{\percent} fidelity even at few-\si{\micro\second} timescales.
We thus demonstrate that transfer-oscillator locking can provide sufficient phase stability for high-fidelity quantum logic manipulation even without pre-stabilization of the slave diode laser.
\end{abstract}


\bibliography{ions} % For OSA, replace this line with the pasted-in content of the .bbl file when submitting.

\noindent
Femtosecond optical frequency combs have dramatically simplified accurate optical-frequency spectroscopy,
allowing the phase-coherent comparison of widely separated optical frequencies~\cite{Cundiff2003}.
Combs based on erbium-doped-fiber lasers~\cite{Newbury2007} are particularly attractive,
as they emit at telecom wavelengths and offer turn-key operation in a reliable and portable instrument~\cite{Schibli2004,Kubina2005,Inaba2006}.
An electro-optic modulator in the laser cavity can provide enough servo bandwidth to phase-lock such a comb to a stable optical reference~\cite{Hudson2005,Zhang2012:comb,Iwakuni2012}.
This can be used to transfer the phase stability of the reference to other wavelengths without any degradation in a band of a few \si{\kilo\hertz} around the carrier~\cite{Akamatsu2012,Inaba2013,Nicolodi2014}.
This bandwidth suffices for precise measurements of optical frequency ratios and for spectroscopy experiments using pulses longer than a few \si{\ms}.
For such experiments, an ensemble of lasers addressing multiple atomic or molecular transitions can share the same carefully optimized reference~\cite{Akamatsu2012,Inaba2013}.

However, experiments involving few-\si{\micro\second} to few-ten-\si{\micro\second} pulses,
such as those in trapped-ion quantum computing and quantum simulation~\cite{Blatt2008,Wineland2013,Schaetz2013},
quantum logic spectroscopy~\cite{Schmidt2005,Rosenband2008},
Rydberg spectroscopy~\cite{Loew2012}
or coherent photoassociation of molecules~\cite{Ospelkaus2008,Danzl2008,Ni2008}
are affected by phase noise at offsets of \SI{100}{\kHz}--\SI{1}{\MHz} from the carrier.
This noise can be substantial even in a phase-locked comb~\cite{Zhang2012:comb},
particularly in fiber combs,
 whose short cavity lifetime increases the fundamental amplified spontaneous emission noise and leads to greater pulse timing jitter than in Ti:Sa oscillators~\cite{Newbury2007,Bao2014}.
Lasers in experiments sensitive to high-frequency phase fluctuations cannot derive their short-term (sub-\si{\ms}) phase stability from the comb,
and are typically locked to separate pre-stabilization cavities,
adding complexity which scales with the number of wavelengths required.

\begin{figure}
  \centering
  \includegraphics{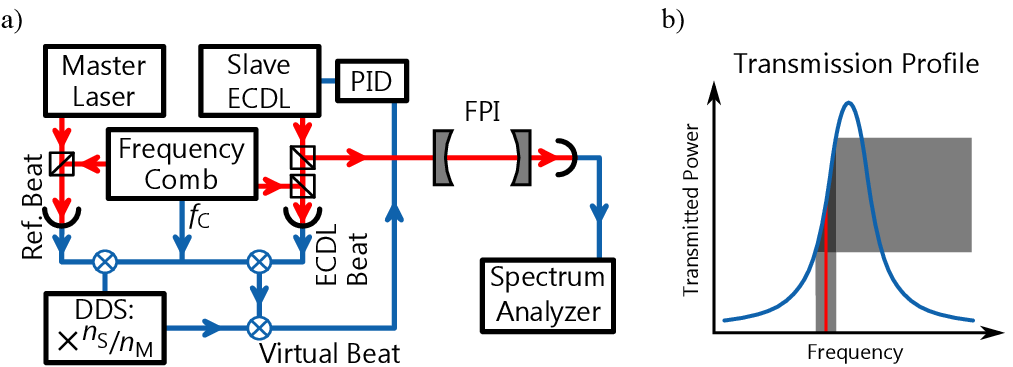}
  \caption{(a) Setup for feed-forward of the repetition-rate noise.
    Electrical signals are in blue, optical signals are in red.
    PID: Proportional-Integral-Derivative feedback controller,
    DDS: Direct Digital Synthesis chip,
    FPI: Fabry-Pérot Interferometer for laser phase-noise measurements.
    b) The FPI converts high-frequency laser frequency noise into intensity noise which can be directly observed with a photodiode and spectrum analyzer.
  }
  \label{fig:setup}
\end{figure}

Here we report the phase-locking of a commercial extended-cavity diode laser (ECDL) operating at \SI{729}{\nm} to a \SI{1542}{\nm} master laser via an erbium-doped fiber comb,
without pre-stabilizing the ECDL to a cavity.
We characterize the noise of the locked comb,
showing that it exceeds the noise of the free-running diode laser beyond \SI{30}{\kHz} from the carrier.
We then implement a high-bandwidth version of the transfer-oscillator feed-forward scheme~\cite{Stenger2002},
measuring the comb noise and using rf electronics to remove it from the error signal in the \SI{729}{\nm} laser's servo loop [Fig.~\ref{fig:setup}(a)].
We demonstrate the practical impact of this enhancement by performing rapid manipulation of the internal state of a trapped laser-cooled \Caion{} ion with the \SI{729}{\nm} laser.
The feed-forward reduces the error probability for a \SI{2.4}{\micro\second} $\pi$ pulse from \SI{10}{\percent} to \SI{1}{\percent},
enabling fast, accurate manipulation of the internal state of the \Caion{}.
This technique ensures the relative phase stability of all lasers locked to a common master,
even at short time-scales,
and thus offers a powerful tool for fast multi-wavelength coherent Raman manipulation.

We use a commercially available frequency comb
(Menlo FC-1500-250-WG with intracavity EOM~\cite{ptbdisclaimer})
based on an \ce{Er^{3+}}-doped fiber amplifier with a repetition rate $\frep = \SI{250.035131}{\MHz}$.
The carrier-envelope offset frequency $\fceo = \SI{40}{\MHz}$ is measured using the comb's built-in $f$--$2f$ interferometer and phase-locked to a reference signal from a hydrogen maser using feedback to the comb's pump current.
The measured $\fceo$ is subtracted from all optical beat frequencies in the system using rf mixers, so that the remaining servo loops are immune to noise in this parameter.
To stabilize the comb's repetition rate, we use the optical beat between a comb tooth with integer index $\nmaster$ and a master laser.
The master laser, maintained by a different research group, is a \SI{1542}{\nm} fiber laser locked to a \SI{10}{\cm}-long ULE reference cavity (a prototype which was subsequently commercialised as the Menlo Systems ORS-1500).
Light at frequency $\fmaster$ is delivered to our laboratory by a noise-cancelled fiber~\cite{Grosche2014}.
After subtraction of $\fceo$,
we obtain a signal at frequency $\nmaster\times\frep - \fmaster$ which we phase-lock to a \SI{57}{\MHz} rf reference,
using feedback to the comb's intracavity EOM and to the piezo that controls the oscillator cavity length~\cite{Hudson2005}.

\begin{figure}
  \centering
  \includegraphics{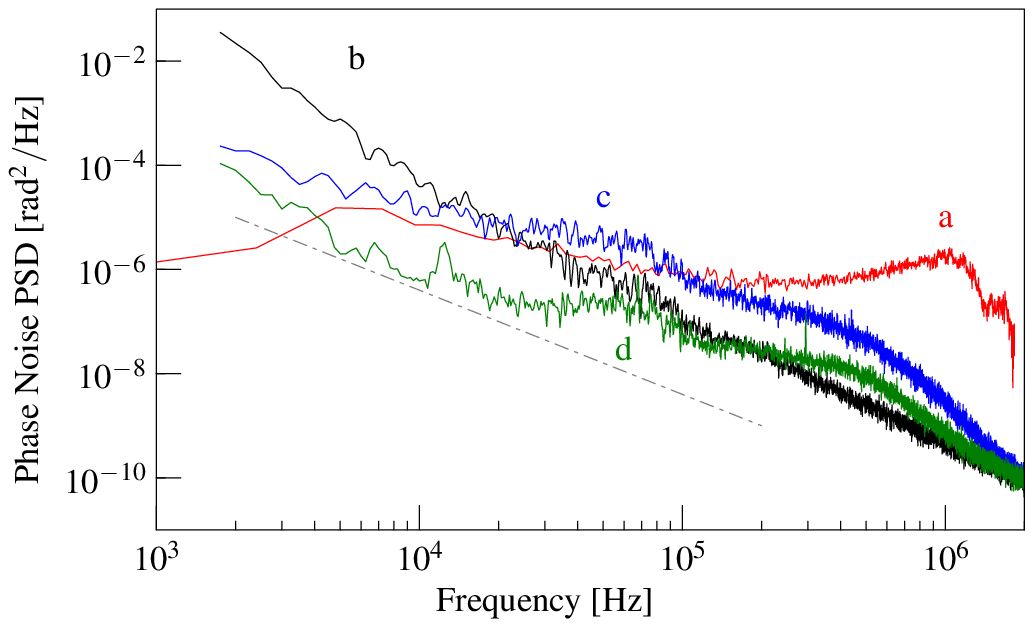}
  \caption{Phase noise power spectral density of
    (a) a comb tooth at \SI{729}{\nm} (after removal of $\fceo$ noise)
    (b) the free-running ECDL, for reference
    (c) the ECDL locked directly to the \SI{729}{\nm} comb tooth
    (d) the ECDL locked to the virtual beat with the \SI{1542}{\nm} master laser.
    The dashed line indicates the measurement noise floor for curves (b--d) due to laser intensity noise.
    Note the additional noise suppression bandwidth of spectrum (d) compared to that of spectrum (c).
  }
  \label{fig:psd}
\end{figure}

Figure~\ref{fig:psd}(a) shows the contribution of residual repetition-rate fluctuations to the phase noise of the comb tooth at \SI{729}{\nm}.
This spectrum was obtained from the in-loop beat between the comb and the \SI{1542}{\nm} master laser,
measured after subtraction of $\fceo$,
by rescaling its phase-noise spectral density by the squared ratio $(\nslave/\nmaster)^2$ of the tooth numbers for the comb teeth at  \SI{729}{\nm} and \SI{1542}{\nm}.
The rescaling step is necessary because fluctuations of $\frep$ introduce larger phase excursions in higher-frequency comb teeth.
The spectrum exhibits servo bumps around \SI{1}{\MHz}
where the servo bandwidth is limited by resonances of the intracavity EOM,
similar to those previously reported by other users of combs of this type~\cite{Zhang2012:comb}.
The spectrum also displays a servo bump around \SI{6}{\kHz},
due to the slow feedback to the mirror piezo,
but most of the residual phase noise is in the region of \SI{100}{\kHz} to \SI{1.3}{\MHz},
the frequency range most relevant for fast (few-\si{\micro\second}-scale) coherent manipulations in experiments with trapped atoms and ions.
Integrating the phase noise out to \SI{2}{\MHz}
we find an rms phase noise of \SI{620}{\milli\radian},
or an energy concentration of \SI{68}{\%} in the carrier.

For comparison, we plot in Fig.~\ref{fig:psd}(b) the phase noise of a free-running \SI{729}{\nm} grating-stabilized diode laser (Toptica TA 100, with a Toptica DLPro seed laser).
We measure the laser's phase noise independently of the comb by observing the fluctuations in the power transmitted through a Fabry-Pérot interferometer
(Toptica FPI-100 modified to lower its finesse and increase its bandwidth to \SI{13.2}{\MHz}).
A slow lock of the interferometer length (bandwidth \SI{<100}{\Hz}) keeps the laser on the slope of the resonance fringe [see Fig.~\ref{fig:setup}(b)], where fast frequency fluctuations are converted into intensity noise which can be directly observed on a photodiode.
Laser intensity noise present before the interferometer sets a noise floor for the measurement, indicated by the grey chain-dotted line in Fig.~\ref{fig:psd}.
Acoustic noise on the resonator length also adds to the noise floor at frequencies below a few \si{\kHz},
but can be neglected for the higher frequencies that interest us here.
Beyond \SI{30}{\kHz} from the carrier, the residual comb noise in Fig.~\ref{fig:psd}(a) exceeds the noise of the free-running ECDL in Fig.~\ref{fig:psd}(b).
When locking the ECDL directly to the comb [Fig.~\ref{fig:psd}(c)],
the servo can only be used to suppress laser noise out to a Fourier frequency of \SI{20}{\kHz}.
Further increasing the servo bandwidth only broadens the diode laser by forcing it to follow the comb's fast phase fluctuations.

We therefore lock the \SI{729}{\nm} ECDL to a virtual beat with the \SI{1542}{\nm} master laser,
in which the effect of repetition-rate noise is suppressed by rf feed-forward electronics.
This so-called transfer-oscillator lock~\cite{Stenger2002},
illustrated in Fig.~\ref{fig:setup}(a),
relies on the fact that repetition-rate noise produces correlated fluctuations of the beat frequencies at \SI{729}{\nm} and \SI{1542}{\nm},
where the absolute frequency deviations of the two beats are related by the ratio $\nslave / \nmaster \approx \num{2.114}$.
After subtraction of $\fceo$,
the beat with the \SI{1542}{\nm} master laser is frequency-multiplied by \num{2.114} and mixed with the \SI{729}{\nm} beat.
The resulting virtual ``transfer beat'' is independent of $\frep$.
This general scheme is commonly used for optical frequency comparisons with fiber combs~\cite{Stenger2002,Nicolodi2014,Godun2014}.
Previously reported implementations only suppress repetition rate noise within a few \si{\kHz} of the carrier,
primarily because of the limited bandwidth of the tracking oscillators used to reject undesired mixer sidebands from the rf signals.
When the transfer beat is used to generate the error signal for a feedback loop,
the resulting restricted servo bandwidth requires pre-stabilization of the slave laser at the sub-\si{\kHz} level~\cite{Nicolodi2014},
and thus an auxiliary reference cavity.
A pure feed-forward scheme offering much greater noise-suppression bandwidth has recently been demonstrated~\cite{Sala2014},
but as this scheme omits the wavelength-dependent rescaling of the comb's repetition-rate fluctuations it is only applicable when the master and slave lasers have similar wavelengths.

To obtain the virtual transfer beat with a delay short enough for high-bandwidth locking of the ECDL,
we avoid the use of tracking oscillators entirely.
Instead, we choose $\fceo$ and $\frep$
(and thus the frequencies of all comb teeth)
such that both beat signals remain well-separated from spurious mirror frequencies generated in the mixers.
The spurs can then be removed using standard rf filters,
keeping the signal of interest well away from the filters' band edge in order to minimize group delay.
In addition, we match the propagation delays in parallel signal paths,
to improve noise cancellation at high Fourier frequencies.
To carry out the necessary frequency scaling,
the $\fceo$-free, filtered and amplified \SI{1542}{\nm} beat note is used as the clock input of a direct digital synthesis chip (DDS, Analog Devices AD9956).
The DDS is programmed to synthesize a frequency of \num{0.114} times its clock input frequency.
Since it emits discrete samples, the output spectrum also contains aliased signals,
including one at \num{2.114} times the clock frequency.
Mixing this spectral component with the beat note from the \SI{729}{\nm} laser produces the virtual transfer beat,
in which the effect of fluctuations in $\frep$ is cancelled up to a \SI{3}{\dB} bandwidth in excess of \SI{1.8}{\MHz}.
We believe this bandwidth to be limited by residual differences in propagation delay between signal paths.
An analog PID controller (Toptica FALC) locks the ECDL to the virtual beat using slow feedback to the ECDL's grating piezo and fast current feedback directly to the laser diode.
The lock has a capture range of approximately \SI{\pm1.5}{\MHz},
limited by the residual offset voltage of the error signal,
and suppresses the phase-noise of the ECDL below the free-running level for frequencies up to \SI{200}{\kHz} [Fig.~\ref{fig:psd}(d)].
Compared to a beat with the comb tooth with subtraction of $\fceo$ alone,
the virtual beat improves the useful locking bandwidth by an order of magnitude.

\begin{figure}
  \centering
  \includegraphics{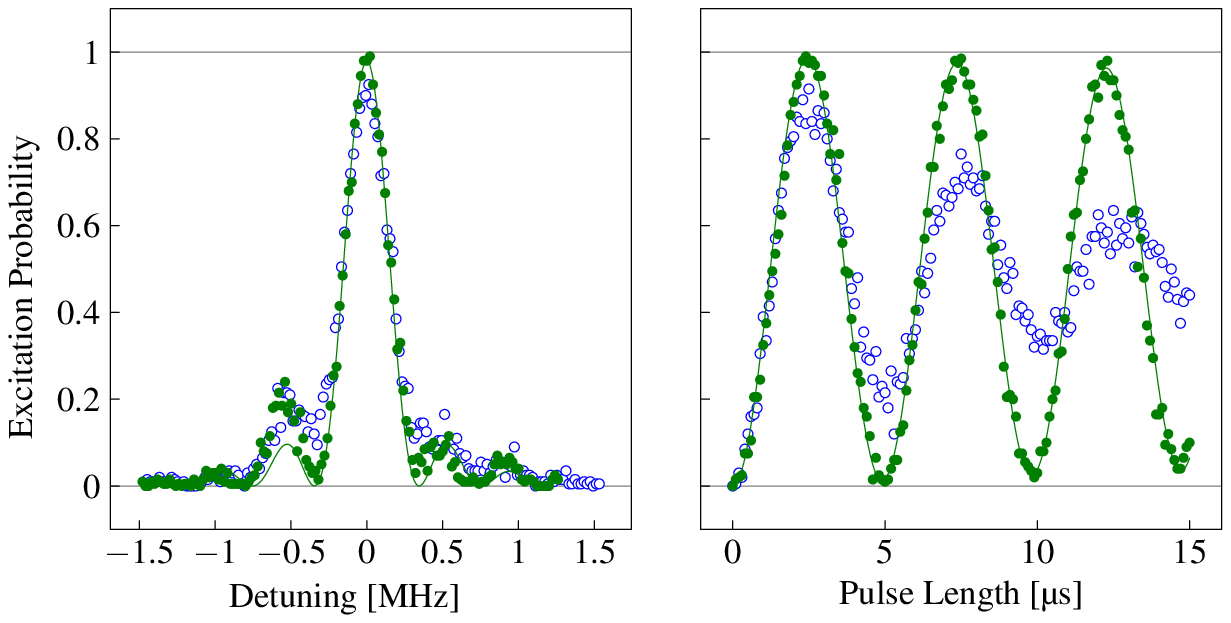}
  \caption{Spectroscopy of the $\trans{\ket*{\Shalf,m=1/2}}{\ket*{\Dfivehalves,m=3/2}}$ transition without (open) and with (filled) feed-forward cancellation of the comb repetition rate noise.
    Inset: Resonant Rabi oscillations.
    With the ECDL locked to the virtual transfer beat we obtain \SI{99}{\%} excitation contrast and a Fourier-limited spectrum for a \SI{2.4}{\micro\second} $\pi$ pulse.
    Without feed-forward the excitation contrast is limited to \SI{90}{\%}.
    Lines are fits to the data with feed-forward of repetition rate noise.}
  \label{fig:spectroscopy}
\end{figure}

With no cavity-stabilized \SI{729}{\nm} laser available for comparison,
we illustrate the effect of high-frequency comb noise,
and the benefits of suppressing it,
by using the ECDL to drive $\pi$ pulses and Rabi oscillations on the narrow
$\trans{\ket*{\Shalf,m=1/2}}{\ket*{\Dfivehalves,m=3/2}}$ transition of a \Caion\ ion confined in a macroscopic linear Paul trap.
Up to \SI{30}{\milli\watt} of power from the ECDL are focused on the ion along the axis of a normal mode with oscillation frequency \SI{3.7}{\MHz} which can be cooled to below \SI{50}{\micro\kelvin} using EIT cooling~\cite{Roos2000}.
A \SI{4.2}{\gauss} magnetic bias field splits the ion's Zeeman levels by \SI{7}{\MHz},
so that  internal and motional levels are well-resolved.
For fast $\pi$ pulses with the ECDL locked directly to the comb,
the best fidelity is obtained by using a loose, low-bandwidth lock to avoid imprinting high-frequency comb noise onto the laser.
With the lock thus optimized, a \SI{2.4}{\micro\second} $\pi$ pulse can have a fidelity as high as \SI{90}{\%},
as shown in Fig.~\ref{fig:spectroscopy} (open symbols).
However, when the ECDL is locked to the virtual transfer beat (filled symbols),
the fidelity is \SI{99}{\%}.
Varying the laser power,
we find that fidelities above \SI{95}{\%} are obtained over a two-order of magnitude range in Rabi frequency,
extending to $\pi$ pulse lengths of \SI{175}{\micro\second},
without any alteration of locking parameters.

We have thus demonstrated that a high-bandwidth transfer-oscillator lock can provide sufficient short-term stability to an otherwise free-running diode laser for coherent internal state manipulation with few-microsecond pulses.
By reducing the need for separate pre-stabilization cavities at each wavelength when phase-locking lasers of different colors,
this technique could simplify experiments involving coherent Raman manipulation,
such as for photoassociation and control of molecules~\cite{Danzl2008}.
The option to use a single transportable reference cavity~\cite{Leibrandt2011,Parker2014} for all lasers in a setup
would also reduce bulk and weight in proposed portable or space-based optical clocks~\cite{Bongs2015} for tests of fundamental physics or geodesy.

\section*{Acknowledgments}
We thank B. Lipphardt for helpful discussions;
T. Legero, G. Grosche and S. Raupach for the use of the \SI{1542}{\nm} master laser and the noise-cancelled fiber link to deliver it to our laboratory;
and A. Koczwara, A. Hoppmann and J. Mielke for technical assistance.
We acknowledge support from ESA and from the DFG through QUEST.
I.D.L. acknowledges a fellowship from the Alexander von Humboldt Foundation.
J.B.W. acknowledges support from the Hannover School for Laser, Optics and Space-Time Research (HALOSTAR) and the German National Academic Foundation (Studienstiftung des deutschen Volkes).
This work was supported by the European Metrology Research Programme (EMRP) in project SIB04.
The EMRP is jointly funded by the EMRP participating countries within EURAMET and the European Union.
This work was financially supported by the State of Lower-Saxony, Hannover, Germany.

\end{document}

%% file: macros.tex
\usepackage[T1]{fontenc}
\usepackage{opex3}
\usepackage[utf8]{inputenc}
\usepackage{mathtools}
\usepackage[version=3]{mhchem}

\usepackage{bm}
\usepackage{siunitx}
\DeclareSIUnit\gauss{G}
\DeclareSIUnit\dBm{dBm}

\DeclarePairedDelimiterX\comm[2]{\lbrack}{\rbrack}{#1,#2}

\DeclarePairedDelimiter\ket{\lvert}{\rangle}
\DeclarePairedDelimiterX\innerp[2]{\langle}{\rangle}{#1\,\delimsize\vert\,#2}
\DeclarePairedDelimiterX\matrixelt[3]{\langle}{\rangle}%
{#1\,\delimsize\vert\,#2\,\delimsize\vert\,#3}

\newcommand*\Caion{\ce{^{40}Ca+}}
\newcommand*\term[3]{\prescript{#1}{}{\mathrm{#2}}_{#3}}
\newcommand*\Shalf{\term{2}{S}{1/2}}

\newcommand*\Dfivehalves{\term{2}{D}{5/2}}
\newcommand*\trans[2]{#1 \leftrightarrow #2}

\newcommand*\fceo{f_\mathrm{C}}
\newcommand*\frep{f_\mathrm{R}}
\newcommand*\fmaster{f_\mathrm{M}}

\newcommand*\nmaster{n_\mathrm{M}}
\newcommand*\nslave{n_\mathrm{S}}

%% file: paper.bbl
\begin{thebibliography}{10}
\bibitem{ptbdisclaimer}
We identify the manufacturer and model of several devices used in this work.
  This is done for reasons of clarity and completeness and does not constitute
  an endorsement by the Physikalisch-Technische Bundesanstalt.
\end{thebibliography}
